\documentclass[aps,pra,reprint,superscriptaddress,nobibnotes]{revtex4-1}

\usepackage{graphicx}
\usepackage{upgreek}
\usepackage{color}
\usepackage{amssymb}
\usepackage{amsmath}
\usepackage{bm}
\renewcommand{\vec}[1]{{\mathbf #1}}

\begin{document}

%\preprint{foo}

\title{Localization transition for light scattering by cold atoms in an external magnetic field}



\author{S.E. Skipetrov}
\email{sergey.skipetrov@lpmmc.cnrs.fr}
\affiliation{Univ. Grenoble Alpes, CNRS, LPMMC, 38000 Grenoble, France}

\date{\today}

\begin{abstract}
We establish a localization phase diagram for light in a random three-dimensional (3D) ensemble of {motionless} two-level atoms with a three-fold degenerate upper level, in a strong static magnetic field. Localized modes appear in a narrow spectral band when the number density of atoms $\rho$ exceeds a critical value $\rho_c \simeq 0.1 k_0^3$, where $k_0$ is the wave number of light in the free space. A critical exponent of the localization transition taking place upon varying the frequency of light at a constant $\rho > \rho_c$ is estimated to be $\nu = 1.57 \pm 0.07$. This classifies the transition as an Anderson localization transition of 3D orthogonal universality class.
\end{abstract}

\maketitle

The search for Anderson localization of light in three-dimensional (3D) disordered media has been an active research direction since the mid-1980s when John \cite{john84} and Anderson \cite{anderson85} independently noticed that light could be localized by strong disorder in a way analogous to electron localization in disordered solids \cite{anderson58}. It was rapidly recognized that optical localization in a dielectric material is difficult to achieve because, on the one hand, of the way in which the disorder enters the optical wave equation (the position-dependent dielectric function $\varepsilon(\vec{r})$ of the material multiplies the second-order time derivative of the electric field) and, on the other hand, of the relatively low values of $\varepsilon$ of available transparent materials at optical frequencies \cite{john91}.
%% As a consequence, the correlation length of disorder (or, equivalently, the typical size of individual scattering particles composing a disordered material) has to be of the order of the optical wavelength in order to achieve strong scattering and ultimately Anderson localization.
Unfortunately, even the materials composed of particles with the largest available dielectric constants, did not allow for an indisputable observation of disorder-induced light localization in 3D thus far \cite{vanderbeek12,sperling16,skip16njp}.

A random spatial arrangement of {motionless} atoms represents an alternative to dielectric media for reaching Anderson localization of light \cite{kaiser99,kaiser09}. Indeed, the coherent backscattering (CBS) of light, considered as a precursor of localization, was observed in cold atomic gases almost 20 years ago \cite{labeyrie99,bidel02,sigwarth04}. However, the vector character of light and the associated dipole-dipole interactions between atoms have been predicted to prevent Anderson localization in atomic systems \cite{skip14,bellando14}. A static external magnetic field partially suppresses the interatomic dipole-dipole interactions and can induce localization of light that is quasiresonant with a $J_g = 0 \to (J_e = 1, m = \pm 1)$ transition ($J_g$ and $J_e$ are the total angular momenta of the atomic ground and excited states, respectively, and $m$ is the magnetic quantum number of the excited state) \cite{skip15}.
It is important to stress that the role played here by the magnetic field is different from that reported in Ref.\ \cite{sigwarth04} where the field enhances the CBS contrast for light scattered by a cloud of Rb atoms. In the latter case, the magnetic field lifts the degeneracy of the atomic \textit{ground} state ($J_g > 0$) and thus suppresses Raman scattering and lengthens the coherence length of light. We consider atoms with a nondegenerate ground state ($J_g = 0$) and no Raman scattering. The magnetic field can have only a negative impact on such interference effects as CBS because the contrast of the latter is already maximum in the absence of the field. Therefore, understanding of Anderson localization in our system cannot be achieved with far-field interference arguments and requires dealing with near-field effects, such as the dipole-dipole interactions.

Although the presence of localized modes in the atomic system subjected to an external magnetic field has been already established in Ref.\ \cite{skip15}, the transition between extended and localized regimes has not been studied yet. This transition takes place in a dense medium that is made strongly anisotropic by the magnetic field, and in the presence of near-field couplings between atoms separated by less than a wavelength in distance. Questions thus arise concerning the nature of this transition: To which extent can it be considered a genuine, disorder-induced Anderson transition? What is its universality class? Does the anisotropy of the atomic medium in a strong magnetic field play any role? It is the purpose of this Letter to provide exhaustive answers to these questions and thereby motivate the experimental work on Anderson localization of light by cold atoms.

An ensemble of $N$ identical two-level atoms (resonance frequency $\omega_0$, $J_g = 0$ for the ground state, $J_e = 1$ for the excited states) at positions $\{ \mathbf{r}_j \}$, $j = 1, \ldots, N$, subjected to a constant external magnetic field $\mathbf{B} \parallel \mathbf{e}_z$ and interacting with a free electromagnetic field, is described by the following Hamiltonian \cite{cohen92, morice95, sigwarth13}:
\begin{eqnarray}
{\hat H} &=& \sum\limits_{j=1}^{N} \sum\limits_{m=-1}^{1} \left(
\hbar \omega_0 + g_e \mu_{B} B m \right) | e_{jm} \rangle
\langle e_{jm}|
\nonumber \\
&+&
\sum\limits_{\bm{\epsilon} \perp \mathbf{k}} \hbar ck
\left( {\hat a}_{\mathbf{k} \bm{\epsilon}}^{\dagger} {\hat a}_{\mathbf{k}\bm{\epsilon}} + \frac12 \right)
- \sum\limits_{j=1}^{N} {\hat{\mathbf{D}}}_j \cdot {\hat{\mathbf{E}}}(\mathbf{r}_j)
\nonumber \\
&+& \frac{1}{2 \varepsilon_0}
\sum\limits_{j \ne n}^{N} {\hat{\mathbf{D}}}_j \cdot {\hat{\mathbf{D}}}_n \delta(\mathbf{r}_j - \mathbf{r}_n).
\label{ham}
\end{eqnarray}
Here we denote the atomic dipole operators by ${\hat{\mathbf{D}}}_j$, the electric displacement vector by $\varepsilon_0 {\hat{\mathbf{E}}}(\mathbf{r})$, the photon creation and annihilation operators corresponding to a mode of the free electromagnetic field having a wave vector  $\mathbf{k}$ and a polarization $\bm{\epsilon}$ by ${\hat a}_{\mathbf{k} \bm{\epsilon}}^{\dagger}$ and ${\hat a}_{\mathbf{k}\bm{\epsilon}}$, respectively.
%% $2\pi\hbar$ is the Planck's constant,
$\mu_{B}$ is the Bohr magneton, and $g_e$ is the Land\'{e} factor of the excited state. As discussed previously \cite{skip15,skip16pra}, the quasimodes of the atomic subsystem can be found as eigenvectors of a $3N \times 3N$ effective Hamiltonian $G$ of the open system of atoms interacting via the electromagnetic field:
\begin{eqnarray}
G_{e_{j m} e_{n m'}} &=& \left(i -  {2 m \Delta} \right) \delta_{e_{j m} e_{n m'}} -
\frac{2}{\hbar \Gamma_0} (1 - \delta_{e_{j m} e_{n m'}})
\nonumber \\
&\times&
\sum\limits_{\mu, \nu}
{d}_{e_{j m} g_j}^{\mu} {d}_{g_n e_{n m'}}^{\nu}
\frac{e^{i k_0 r_{jn}}}{r_{jn}^3}
\nonumber
\\
&\times& \left\{
\vphantom{\frac{r_{jn}^{\mu} r_{jn}^{\nu}}{r_{jn}^2}}
 \delta_{\mu \nu}
\left[ 1 - i k_0 r_{jn} - (k_0 r_{jn})^2 \right]
\right.
\nonumber \\
&-&\left. \frac{r_{jn}^{\mu} r_{jn}^{\nu}}{r_{jn}^2}
\left[3 - 3 i k_0 r_{jn} - (k_0 r_{jn})^2 \right]
\right\},
\label{green}
\end{eqnarray}
where $k_0 = \omega_0/c$, $\Delta = g_e \mu_{B} B/\hbar\Gamma_0$ is the Zeeman shift in units of the spontaneous decay rate $\Gamma_0$, $\vec{d}_{e_{j m} g_j} = \langle J_{e} m|{\hat{\mathbf{D}}}_j | J_{g} 0 \rangle$, and $\vec{r}_{jn} = \vec{r}_j - \vec{r}_n$. The complex eigenvalues $\Lambda_n$ of the matrix $G$ yield eigenfrequencies $\omega_n = \omega_0 - (\Gamma_0/2) \mathrm{Re} \Lambda_n$ and decay rates $\Gamma_n/2 = (\Gamma_0/2) \mathrm{Im} \Lambda_n$ of quasimodes. From here on, we consider atoms that are randomly distributed in a ball of radius $R$ and volume $V$ with an average density $\rho = N/V$.

In a strong magnetic field, the eigenvalues $\Lambda_n$ split in three groups corresponding to transitions between the ground state and one of the three Zeeman sublevels ($m = 0, \pm 1$) which now have different frequencies $\omega_m = \omega_0 + m \Gamma_0 \Delta$ \cite{skip15}. Each group occupies a roughly circular area of radius $b_0 \sim R/\ell_0$ on the complex plane \cite{skip11,sm}. Here $\ell_0 \sim k_0^2/\rho$ is the on-resonance scattering mean-free path computed in the independent-scattering approximation (ISA) and $b_0$ is the on-resonance optical thickness. If the distance between eigenvalue groups on the complex plane $2\Delta$ is much larger than $2 b_0$, each group can be found independently
%% from the two others
by diagonalizing an $N \times N$ matrix ${\cal G}^{(m)}$. For $m = \pm 1$, we find \cite{sm}:
\begin{eqnarray}
{\cal G}^{(\pm 1)}_{jn} &=&
(i \mp 2 \Delta {) \delta_{jn}} + (1-\delta_{jn}) \frac{3}{2} \frac{e^{i k_0 r_{jn}}}{k_0 r_{jn}}
\nonumber \\
&\times& \left[ P(i k_0 r_{jn}) + Q(i k_0 r_{jn}) \frac{\sin^2 \theta_{jn}}{2} \right],
\label{gfull}
\end{eqnarray}
where $P(x) = 1 - 1/x + 1/x^2$ and $Q(x) = -1 + 3/x -3/x^2$.
Note that Eq.\ (\ref{gfull}) still contains divergent near-field terms $\propto 1/r_{jn}^3$ associated with dipole-dipole interactions between atoms, but their magnitude is partially suppressed with respect to the case of $\vec{B} = 0$ \cite{skip14}. The effective anisotropy of the atomic medium in a strong magnetic field, which is not obvious from Eq.\ (\ref{green}), now becomes evident because Eq.\ (\ref{gfull}) contains an explicit dependence on the angle $\theta_{jn}$ between $\vec{r}_{jn}$ and $\vec{B}$. A comparison of eigenvalues of the matrices (\ref{green}) and (\ref{gfull}) and a discussion of the condition of validity $\Delta \gg b_0$ of Eq.\ (\ref{gfull}) can be found in the Supplemental Material \cite{sm}.

\begin{figure}[t]
\includegraphics[width=0.99\columnwidth]{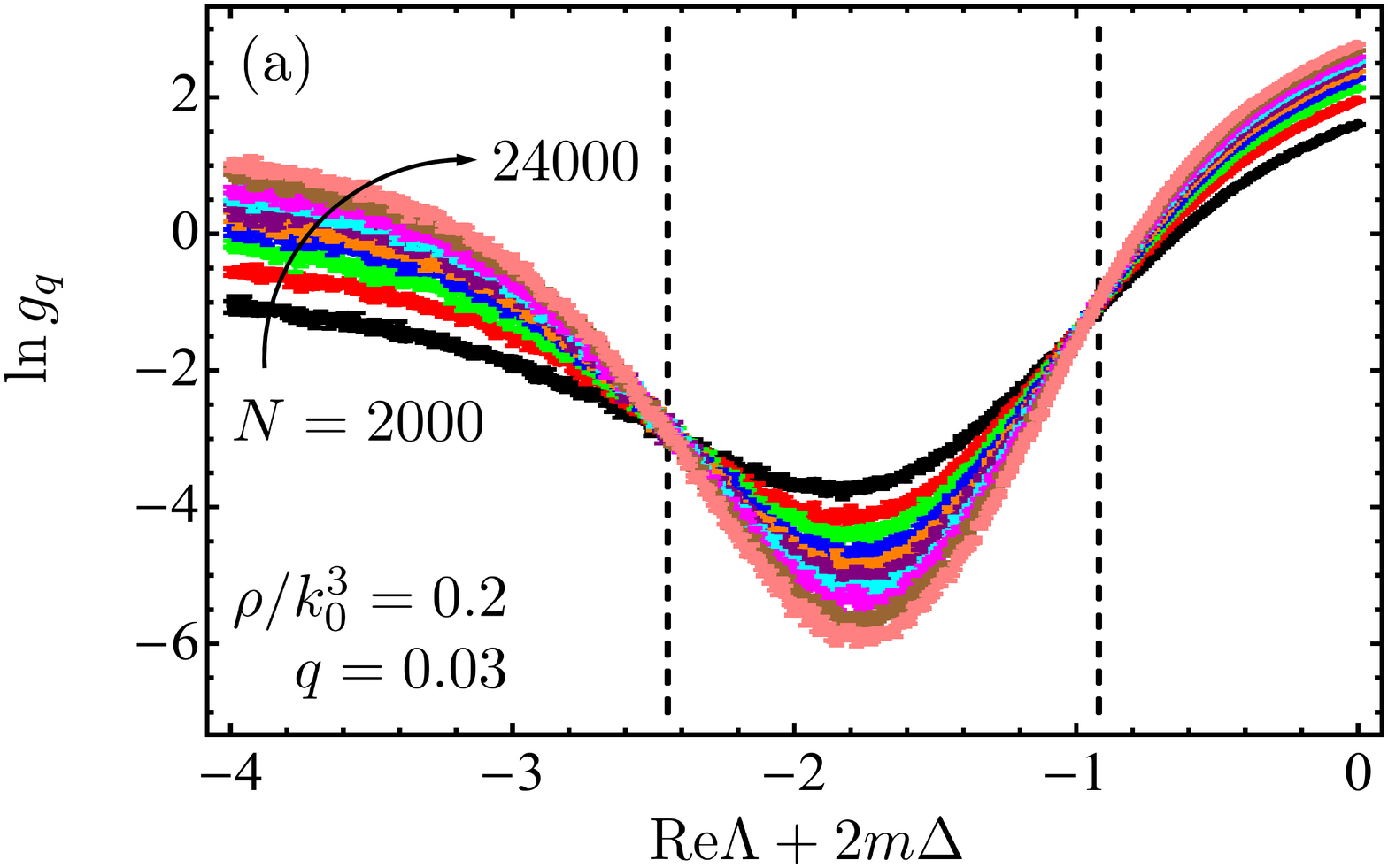}\\
\vspace{-7mm}
\includegraphics[width=0.99\columnwidth]{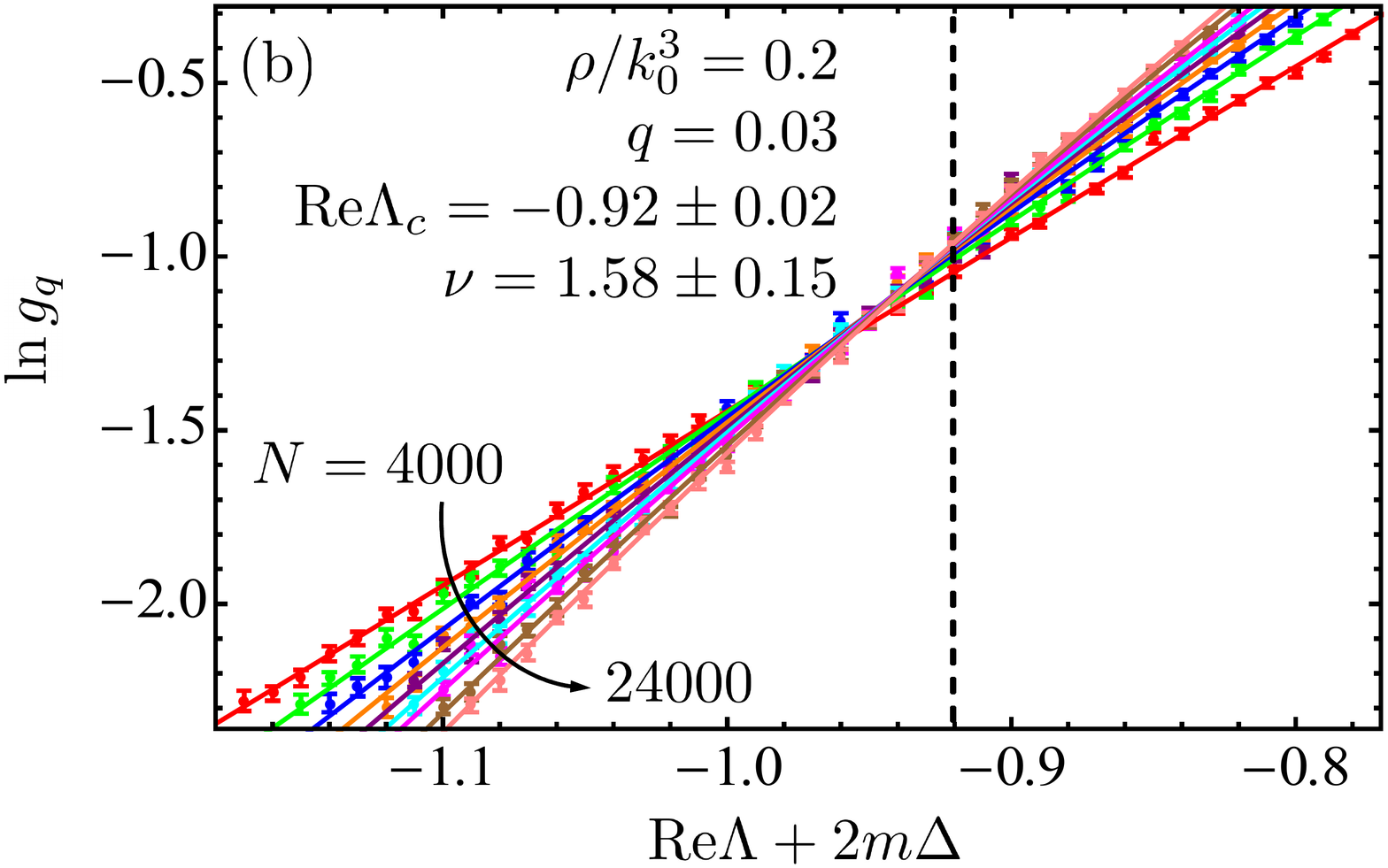}
\vspace{-12mm}
\caption{(a) Third percentile ($q= 0.03$) of the logarithm of Thouless conductance $g$ as a function of frequency at a fixed number density of atoms $\rho/k_0^3 = 0.2$ for 10 different sizes of the atomic cloud: $N = 2000$, 4000, 6000, 8000, 10000, 12000, 14000, 16000, 20000 and 24000 (symbols with error bars). At least $2 \times 10^7$ eigenvalues were calculated for each $N$ to compute the percentile. Vertical dashed lines indicate the mobility edges. (b) Fits (solid lines) of Eqs.\ (\ref{lngqseries}) and (\ref{useries})
%% with $m_1 = 1$, $n_1 = 2$, $m_2 = n_2 = 1$
to numerical data for $N \geq 4000$ (symbols with error bars) around one of the  mobility edges. The best-fit values of the mobility edge $\mathrm{Re} \Lambda_c$ and of the critical exponent $\nu$ are given on the graph.}
\label{fig_fit}
\end{figure}

We will use Eq.\ (\ref{gfull}) to study the localization transition for light that is quasiresonant with the transition between the ground state and one of the excited states corresponding to $m = \pm 1$ (there is no localization transition for $m = 0$ \cite{skip15}). To identify the critical points (mobility edges), we use the approach developed in Ref.\ \cite{skip16prb} for scalar waves. In brief, at a fixed (and sufficiently high) density $\rho$ and for a set of different atom numbers $N$, we compute the eigenvalues $\Lambda_n$ of the matrix (\ref{gfull}) for an ensemble of random atomic configurations $\{ \vec{r}_j \}$ and then estimate the probability density $p(\ln g; \mathrm{Re} \Lambda, N)$ of the logarithm of the Thouless conductance $g = \mathrm{Im} \Lambda_n/\langle \mathrm{Re} \Lambda_n - \mathrm{Re} \Lambda_{n-1} \rangle$, where the angular brackets denote ensemble averaging. The small-$g$ part of $p(\ln g; \mathrm{Re} \Lambda, N)$ becomes independent of $N$ at the critical points $\mathrm{Re} \Lambda_c$. Instead of working with $p(\ln g; \mathrm{Re} \Lambda, N)$, it is more convenient to analyze its low-rank ($q \leq 0.05$) percentiles $\ln g_q$ defined by $q = \int_{-\infty}^{\ln g_q} p(\ln g; \mathrm{Re} \Lambda, N) d(\ln g)$. Figure \ref{fig_fit}(a) shows the third percentile ($q = 0.03$) as a function of $\mathrm{Re} \Lambda$ for different $N$. The crossing points of lines corresponding to different $N$ provide approximate positions of mobility edges $\mathrm{Re} \Lambda_c$ shown by vertical dashed lines.

\begin{figure}[t]
\includegraphics[width=0.99\columnwidth]{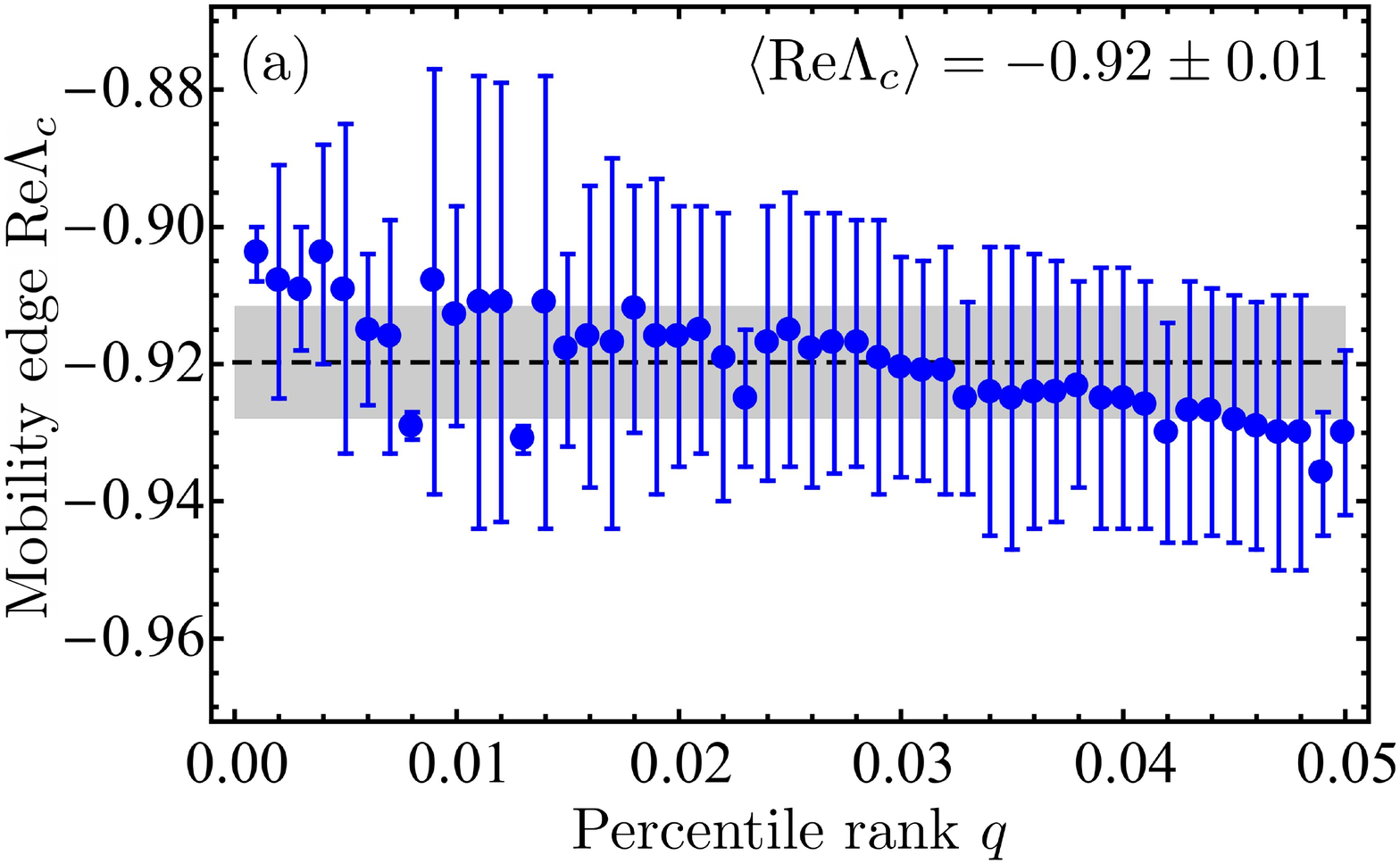}\\
\vspace{-7mm}
\includegraphics[width=0.99\columnwidth]{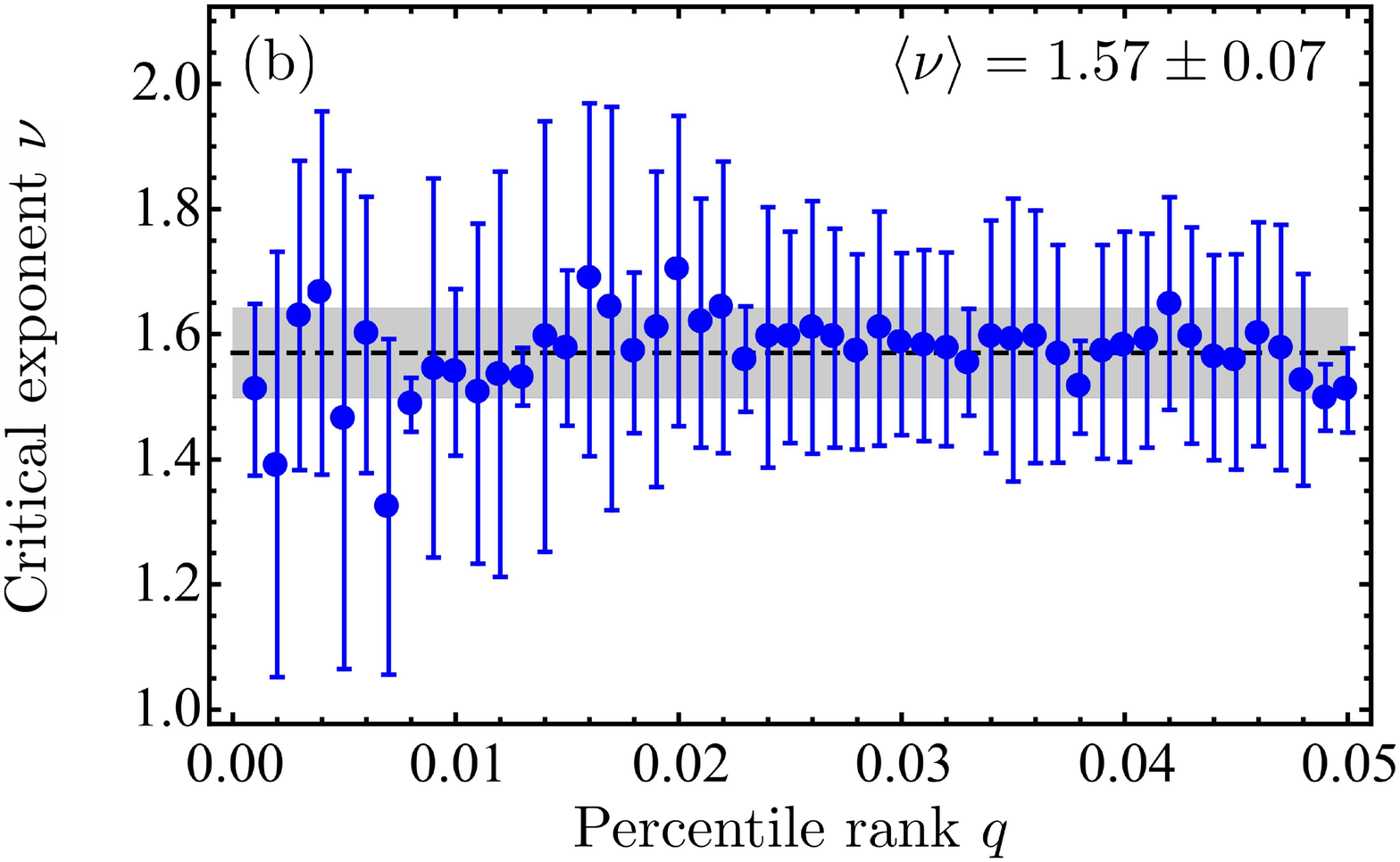}
\vspace{-12mm}
\caption{Best-fit values of the mobility edge $\mathrm{Re} \Lambda_c$ (a) and of the critical exponent $\nu$ (b) as functions of the rank $q$ of the analyzed percentile $\ln g_q$. Fits were performed for the data corresponding to $\rho/k_0^3 = 0.2$ and $\ln g_q$ within $\pm 1$ of an estimated crossing point of curves obtained for different $N$.
%% with $m_1 = 1$, $n_1 = 2$, $m_2 = n_2 = 1$.
The dashed solid lines show average values of $\mathrm{Re} \Lambda_c$ and $\nu$, respectively, with the values of the latter printed on the graphs. Grey areas visualize the errors of the averages.}
\label{fig_best}
\end{figure}

The finite-size scaling analysis of the localization transition consists in fitting the numerical data for $\ln g_q$ near a critical point by polynomials \cite{skip16prb,slevin14}:
\begin{eqnarray}
\ln g_q &=& \sum\limits_{j_1=0}^{n_1} \sum\limits_{j_2=0}^{n_2} a_{j_1 j_2}
u_1(w)^{j_1}
u_2(w)^{j_2}
(k_0 R)^{j_1/\nu + j_2 y},\;\;\;\;\;
\label{lngqseries}
\\
u_i(w) &=& \sum\limits_{j=0}^{m_i} b_{ij} w^j,
\label{useries}
\end{eqnarray}
where $w = (\mathrm{Re} \Lambda - \mathrm{Re} \Lambda_c)/\mathrm{Re} \Lambda_c$, $\nu$ is a critical exponent of the localization transition, and $y < 0$ is an irrelevant exponent accounting for {deviations from the single-parameter scaling}. $m_1 = 1$, $n_1 = 2$, $m_2 = n_2 = 1$ in Eqs.\ (\ref{lngqseries}) and (\ref{useries}) are the minimum values that yield fits of acceptable quality and, at the same time, give consistent values of best-fit $\mathrm{Re} \Lambda_c$ and $\nu$ for all $q$ from 0.001 to 0.05. An example of fit is shown in Fig.\ \ref{fig_fit}(b) whereas Fig.\ \ref{fig_best} shows the best-fit values of the mobility edge, corresponding to $\mathrm{Re} \Lambda_c + 2 m \Delta \simeq -1$ in Fig.\ \ref{fig_fit}(a), and of the critical exponent $\nu$ as functions of the rank $q$ of the considered percentile for $\rho/k_0^3 = 0.2$. The analysis of the second mobility edge [$\mathrm{Re} \Lambda_c + 2 m \Delta \simeq -2.4$ in Fig.\ \ref{fig_fit}(a)] is complicated by a stronger noise in the numerical data and does not yield reliable estimations of $\nu$ with an acceptable precision.

The best estimate of the critical exponent $\langle \nu \rangle = 1.57 \pm 0.07$ obtained by averaging results obtained for all $q = 0.001$--0.05 [see Fig.\ \ref{fig_best}(b)], is consistent with the value expected for the Anderson transition of the 3D orthogonal universality class, typical for spinless time-reversal (TR) invariant systems \cite{slevin14,evers08}.
The Hamiltonian (\ref{ham}) is formally invariant under TR of the whole system ``light + atoms + the magnet creating the magnetic field $\mathbf{B}$'' (remember that $\mathbf{B}$ changes sign upon time reversal) \cite{tiggelen98}. However, for a constant $\mathbf{B}$ that we consider, the subsystem ``light + atoms'' is not TR invariant and one might expect the localization transition to belong to the unitary universality class and have a different critical exponent \cite{evers08,slevin96}.
%% The situation with reciprocity is more subtle. On the one hand, if the initial excitation is in the electromagnetic subsystem and is carried by a photon with a wave vector $\mathbf{k}$ and a polarization $\bm{\epsilon}$, it is easy to check that the reciprocity is broken because the amplitude of a wave scattered from the initial state ($\mathbf{k}$, $\bm{\epsilon}$) into a final state ($\mathbf{k}'$, $\bm{\epsilon}'$) is different from the amplitude of a wave scattered from ($-\mathbf{k}'$, $\bm{\epsilon}'$) to ($-\mathbf{k}$, $\bm{\epsilon}$).
To resolve this apparent contradiction, let us consider the transfer of an excitation from an atom $n$ to a distant atom $j$ (see Sec. I of Supplemental Material \cite{sm} for details). The transfer is operated by photons of different helicities $\epsilon = \pm 1$ with {probability} amplitudes $A_{jn}^{(\epsilon)}$. It is not {TR} invariant because the transfer of an excitation from the atom $j$ back to the atom $n$ {by a photon with the same helicity have a different probability amplitude:} $A_{nj}^{(\epsilon)} \ne A_{jn}^{(\epsilon)}$. However, this exchange of photons have an additional symmetry imposing $A_{nj}^{(\epsilon)} = A_{jn}^{(-\epsilon)}$. In other words, the photons of positive (negative) helicity play the same role in the excitation transfer from one atom to another as the photons of negative (positive) helicity do for the transfer in the opposite direction. As a result,
{the exchange of excitations between atoms become TR invariant}, and the localization transition in {the ensemble of atoms} belongs to the orthogonal universality class  \cite{note1}.

The anisotropy induced in the atomic medium by the external magnetic field may play a role in determining the mobility edges \cite{abrikosov94,kaas08, milde00}, but apparently does not modify the universality class of the localization transition, in agreement with {both} previous theoretical results for the anisotropic Anderson model \cite{milde00} and experiments in cold-atom systems \cite{lopez12}.

\begin{figure}[t]
\includegraphics[width=0.99\columnwidth]{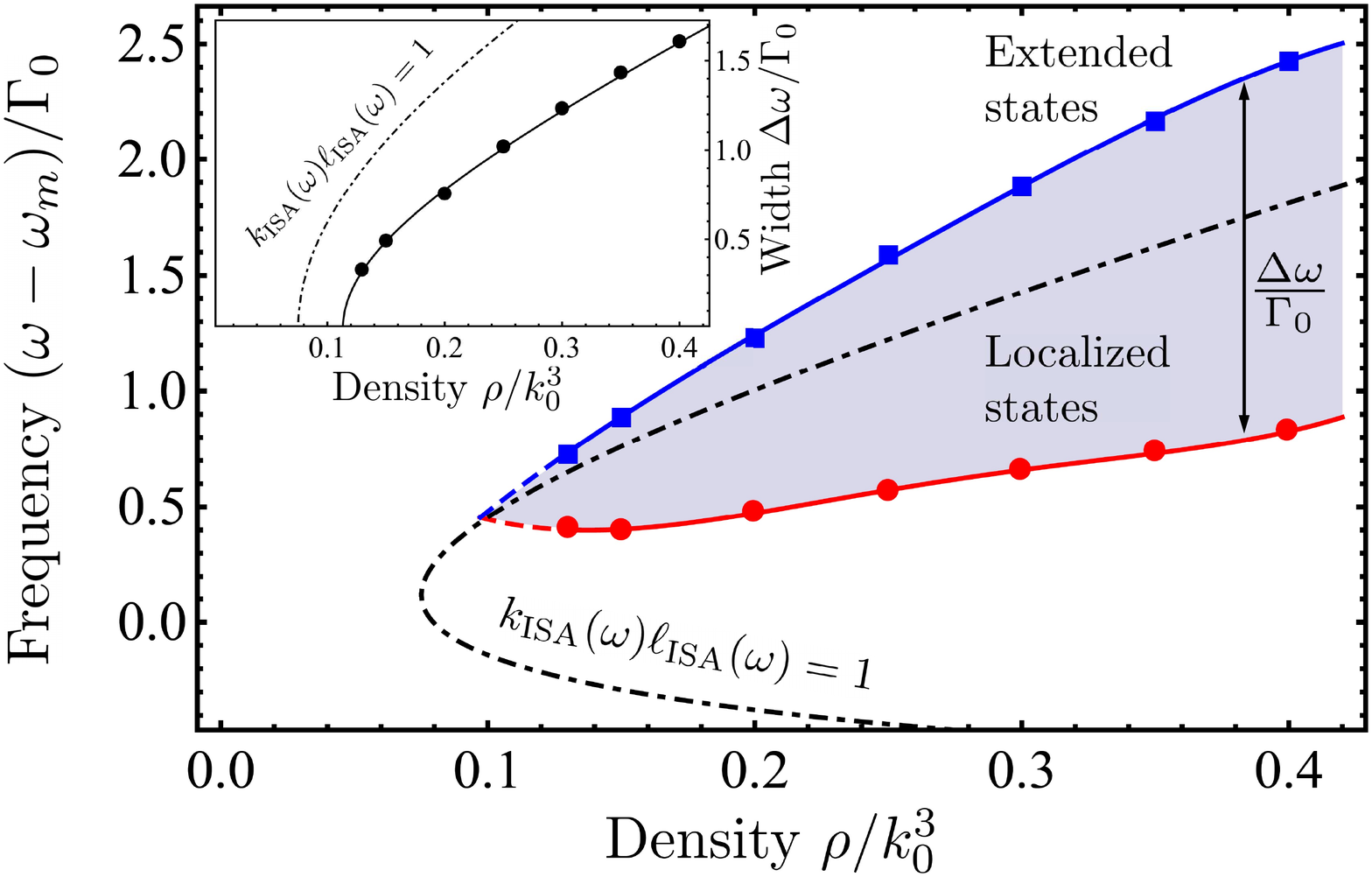}\\
\vspace{-7mm}
\caption{Localization phase diagram of an ensemble of two-level atoms in a strong magnetic field. The frequency on the vertical axis is measured from $\omega_m = \omega_0 + m \Gamma_0 \Delta$ (for $m = \pm 1$) in units of $\Gamma_0$. Symbols are mobility edges calculated from at least $10^7$ ($5.5 \times 10^6$) eigenvalues for $N = 8000$ (16000) at every $\rho$; lines are polynomial fits. The crossing point of the latter $\rho_c/k_0^3 \simeq 0.1$ is an estimation of the absolute localization threshold.
Inset: the width of the frequency band of localized states (circles) fitted by Eq.\ (\ref{width}) (solid line) with $\rho_c/k_0^3 \simeq 0.11$ and $\rho^*/k_0^3 \simeq 0.51$. The dash-dotted lines in both the main plot and the inset are obtained from the IR criterion in the ISA \cite{sm}.}
\label{fig_phase_diagram}
\end{figure}

The analysis performed above for a single atomic density $\rho/k_0^3 = 0.2$ can be repeated for other densities as well. The calculation of $\nu$ is very computer-time consuming but the mobility edges can be estimated from the results for a smaller number of random configurations $\{ \vec{r}_j \}$  and only two different $N$, see Fig.\ \ref{fig_phase_diagram}. For $0.1 \lesssim \rho/k_0^3 \lesssim 0.12$ the mobility edges are too close to be clearly distinguishable, and they disappear for $\rho/k_0^3 < \rho_c/k_0^3 \simeq 0.1$
%% (i.e., lines $\ln g_q$ corresponding to different $N$ do not cross).
The latter value also follows from polynomial fits in Fig.\ \ref{fig_phase_diagram} as a minimum density at which localized states appear. It is slightly larger than $\rho_c/k_0^3 \simeq 0.08$ identified as the absolute localization threshold for scalar waves \cite{skip18}. In our opinion, this difference reflects the residual dipole-dipole interactions which are partially suppressed but not fully eliminated by the magnetic field. As a consequence, Anderson localization requires a higher scatterer density and thus is more difficult to reach for light scattered by atoms in a magnetic field than for scalar waves. In addition to this, the region of localized states in the phase diagram of Fig.\ \ref{fig_phase_diagram} is significantly shifted upwards with respect to its counterpart for scalar waves \cite{skip18}, its shape is modified and its width is reduced.

Some features of Fig.\ \ref{fig_phase_diagram} can be qualitatively understood based on the Ioffe-Regel (IR) criterion of Anderson localization $k \ell = 1$ evaluated in the ISA.
%% Here $k(\omega)$ is the effective wave number and $\ell(\omega)$ is the scattering mean free path. Assuming that light is quasiresonant with the transition $J_g = 0$ $\to$ ($J_e = 1$, $m = \pm 1$) having a resonance frequency $\omega_m$ and a width $\Gamma_0$, and
Calculating the effective wave number $k(\omega)$ and the scattering mean free path $\ell(\omega)$ in the scalar approximation and in the lowest order in density $\rho$ {for an atomic resonance at $\omega = \omega_m$}, we obtain the dashed-dotted lines in Fig.\ \ref{fig_phase_diagram} \cite{sm}. The ISA yields a correct order of magnitude for the minimum density $\rho_c$ needed to reach localization and rightly predicts that the band of localized states is blue shifted with respect to $\omega_m$, {although} it largely underestimates the {magnitude} of the shift. The most obvious failure of the ISA in Fig.\ \ref{fig_phase_diagram} is its complete incapacity to describe the low-frequency mobility edge. As a consequence, also the width $\Delta \omega$ of the spectral band in which Anderson localization takes place is overestimated. The ISA yields a compact expression for it \cite{sm}:
\begin{eqnarray}
&&\frac{\Delta \omega}{\Gamma_0} = \frac{\pi}{k_0^3} \sqrt{\left(  \rho - \rho_c  \right)
\left(  \rho + \rho^*  \right)},
\label{width}
\end{eqnarray}
where $\rho_c = k_0^3 (\sqrt{5}-2)/\pi$ and $\rho^* = k_0^3 (\sqrt{5}+2)/\pi$. This equation is shown in the inset of Fig.\ \ref{fig_phase_diagram} by a dash-dotted line. Although Eq.\ (\ref{width}) does not describe our numerical results shown by symbols, it can provide a good fit to them if we treat $\rho_c$ and $\rho^*$ as free fit parameters (solid line in the inset of Fig.\ \ref{fig_phase_diagram}).

It is important to {keep in mind} that in this work we take two limits in a well-defined order: first  $B \to \infty$ and then $N \to \infty$, ensuring that the condition $\Delta \gg b_0$ needed to justify Eq.\ (\ref{gfull}) is always obeyed.
%% The first limit allows us to reduce the dimension of the considered random matrix $G$ and to work with a smaller matrix ${\cal G}^{(\pm 1)}$, whereas the second one is implied by the finite-size scaling procedure.
Changing the order of limits would require working with the full $3N \times 3N$ matrix $G$ and might modify the results. In an experiment with cold atoms,
%% neither $B \to \infty$ nor $N \to \infty$ is realizable, but it is nevertheles
it should be possible to achieve sufficiently strong fields $B$ to decouple transitions with different magnetic quantum numbers $m$ for a given $N$ (see Ref.\ \cite{skip16pra} for a discussion of a possible experiment). The optical localization transition can be studied by observing the time-dependent fluorescence of a dense atomic system after an excitation by a pulse \cite{skip16pra}. The fluorescence is expected to slow down when the frequency and the polarization of the exciting pulse correspond to those of localized quasimodes. Spectral analysis of the signal should allow for identification of frequencies $\omega_n$ and decay rates $\Gamma_n$ of quasimodes as it has been already done in quasi-1D  microwave experiments \cite{wang11}, allowing for calculation of Thouless conductance $g(\omega)$ for each atomic configuration. Using ``artificial atoms'' (quantum dots) may have an advantage of obviating the Doppler and recoil effects inherent for atomic systems \cite{cohen92}.

In conclusion, we established the localization phase diagram for light propagating in a {dense} random 3D ensemble of two-level atoms subjected to a strong magnetic field, and {estimated} the critical exponent $\nu$ of the localization transition taking place upon varying its frequency.
%% In conclusion, we established the phase diagram and {estimated} the critical exponent $\nu$ of the localization transition taking place upon varying the frequency of quasiresonant light propagating in a {dense} random 3D ensemble of two-level atoms subjected to a strong magnetic field.
The value $\nu = 1.57 \pm 0.07$ obtained from a finite-size scaling analysis, indicates that the transition is an Anderson transition of 3D orthogonal universality class, despite the broken time-reversal symmetry of the system ``light $+$ atoms'' which might have changed its universality class into the unitary one, but is compensated by a symmetry between photons of opposite helicities propagating in opposite directions. The transition frequencies are blue-shifted with respect to the atomic resonances $\omega_0 \pm \Gamma_0 \Delta$ and the minimum number density of atoms required to reach the transition is $\rho_c \simeq 0.1 k_0^{3}$. These features of the phase diagram are qualitatively reproduced by the IR criterion of localization evaluated in the ISA. However, the latter criterion fails to provide quantitatively accurate results. The anisotropy induced in the atomic medium by the external magnetic field does not modify the universality class of the localization transition.

%% \begin{acknowledgments}
This work was funded by the Agence Nationale de la Recherche (grant ANR-14-CE26-0032 LOVE). All the computations presented in this Letter were performed using the Froggy platform of the CIMENT infrastructure ({\tt https://ciment.ujf-grenoble.fr}), which is supported by the Rhone-Alpes region (grant CPER07\verb!_!13 CIRA) and the Equip@Meso project (reference ANR-10-EQPX-29-01) of the programme Investissements d'Avenir supervised by the Agence Nationale de la Recherche.
%% \end{acknowledgments}

\renewcommand{\vec}[1]{{\mathbf #1}}
\renewcommand{\thefigure}{S\arabic{figure}}
\renewcommand{\theequation}{S\arabic{equation}}

\bibliographystyle{apsrev4-1}
\renewcommand*{\citenumfont}[1]{S#1}
\renewcommand*{\bibnumfmt}[1]{[S#1]}

\setcounter{equation}{0}
\setcounter{figure}{0}

\newpage

\onecolumngrid

\begin{center}
\Large{\bf{Supplemental Material}}
\end{center}

\begin{center}
\parbox{0.75\textwidth}{\small Here we present a heuristic derivation and a verification of validity of the model (3) used in the main text, and a derivation of the localization phase diagram from the Ioffe-Regel criterion in the independent-scattering approximation.}
\end{center}

\twocolumngrid

\section{Heuristic derivation of the effective scalar Hamiltonian for atoms in a strong magnetic field}
\label{deriv}

Assume that the atoms are in a strong magnetic field $\Delta \to \infty$ and consider the probability amplitude for a transfer of an excitation from an atom $n$ (located at $\vec{r}_n$) in an excited state with, say, $m = -1$, to an atom $j$ (located at $\vec{r}_j$) initially in the ground state. For the purpose of argument, assume that the atoms are far from each other: $k_0 r_{jn} \gg 1$, $\vec{r}_{jn} = \vec{r}_{j} - \vec{r}_{n}$. The atom $n$ emits a photon having a frequency around $\omega_{-1} = \omega_0 - \Gamma_0 \Delta$, a wave vector $\mathbf{k} \parallel \mathbf{r}_{jn}$, and a helicity $\bm{\epsilon} \perp \mathbf{k}$ with a probability amplitude proportional to a scalar product $\bm{\epsilon}^* \cdot \mathbf{d}_{g_n e_{nm}} \propto (1 - \epsilon \cos\theta_{jn})\exp(-i \varphi_{jn})$, where $\epsilon = \pm 1$ for right and left helicities, respectively, and $\mathbf{r}_{jn} = \{ r_{jn}, \theta_{jn}, \varphi_{jn} \}$. The photon will reach the atom $j$ with a probability amplitude $\propto \exp(i k_0 r_{jn})/k_0 r_{jn}$ describing the propagation of a spherical wave, and will be absorbed on the transition $J_g = 0 \to (J_e = 1, m = -1)$ of this atom with a probability amplitude proportional to $\bm{\epsilon} \cdot \mathbf{d}_{e_{jm} g_j} \propto (1 - \epsilon \cos\theta_{jn})\exp(i \varphi_{jn})$. The two other transitions corresponding to $m = 0$ and $m = +1$ cannot be excited because their resonant frequencies are too different from $\omega_{-1}$. {For a given helicity $\epsilon$,} the probability amplitude {$A_{jn}^{(\epsilon)}$} of the excitation transfer is obtained by multiplying the probability amplitudes of emission by the atom $n$, propagation from $\mathbf{r}_n$ to $\mathbf{r}_j$, and absorbtion by the atom $j \ne n$:
{
\begin{eqnarray}
A_{jn}^{(\epsilon)} &\propto& \frac{e^{i k_0 r_{jn}}}{k_0 r_{jn}} (1 - \epsilon\cos\theta_{jn})^2.
\label{ajn}
\end{eqnarray}
Summation over the two helicities yields the total probability amplitude of the excitation transfer:
\begin{eqnarray}
{\cal G}^{(-1)}_{jn} &=& \sum_{\epsilon=\pm 1} A_{jn}^{(\epsilon)}
\propto \frac{e^{i k_0 r_{jn}}}{k_0 r_{jn}} \left( 1- \frac12 \sin^2 \theta_{jn} \right).
\label{gprop1sm}
\end{eqnarray}
}
Repeating all the reasonings for $m = +1$, we obtain exactly the same equation for ${\cal G}^{(+1)}_{jn}$, whereas the result for ${\cal G}^{(0)}_{jn}$ is different:
\begin{eqnarray}
{\cal G}^{(0)}_{jn} &\propto& \frac{e^{i k_0 r_{jn}}}{k_0 r_{jn}} \left( 1 - \cos^2 \theta_{jn} \right).
\label{gprop0sm}
\end{eqnarray}
Extending the above analysis to arbitrary $r_{jn}$ by properly including near-field dipole-dipole interactions between the atoms and keeping trace of all numerical factors yields
\begin{eqnarray}
{\cal G}^{(\pm 1)}_{jn} &=&
(i \mp 2 \Delta {) \delta_{jn}} + (1-\delta_{jn}) \frac{3}{2} \frac{e^{i k_0 r_{jn}}}{k_0 r_{jn}}
\nonumber \\
&\times& \left[ P(i k_0 r_{jn}) + Q(i k_0 r_{jn}) \frac{\sin^2 \theta_{jn}}{2} \right],
\label{g1sm} \\
{\cal G}^{(0)}_{jn} &=&
i \delta_{jn} + (1-\delta_{jn}) \frac{3}{2} \frac{e^{i k_0 r_{jn}}}{k_0 r_{jn}}
\nonumber \\
&\times& \left[ P(i k_0 r_{jn}) + Q(i k_0 r_{jn}) \cos^2 \theta_{jn} \right],
\label{g0sm}
\end{eqnarray}
where $P(x) = 1 - 1/x + 1/x^2$ and $Q(x) = -1 + 3/x -3/x^2$.
Equation (\ref{g1sm}) is Eq.\ (3) of the main text.

\section{Study of validity of the scalar Hamiltonian (3)}
\label{test}

\begin{figure*}
\includegraphics[width=0.99\textwidth]{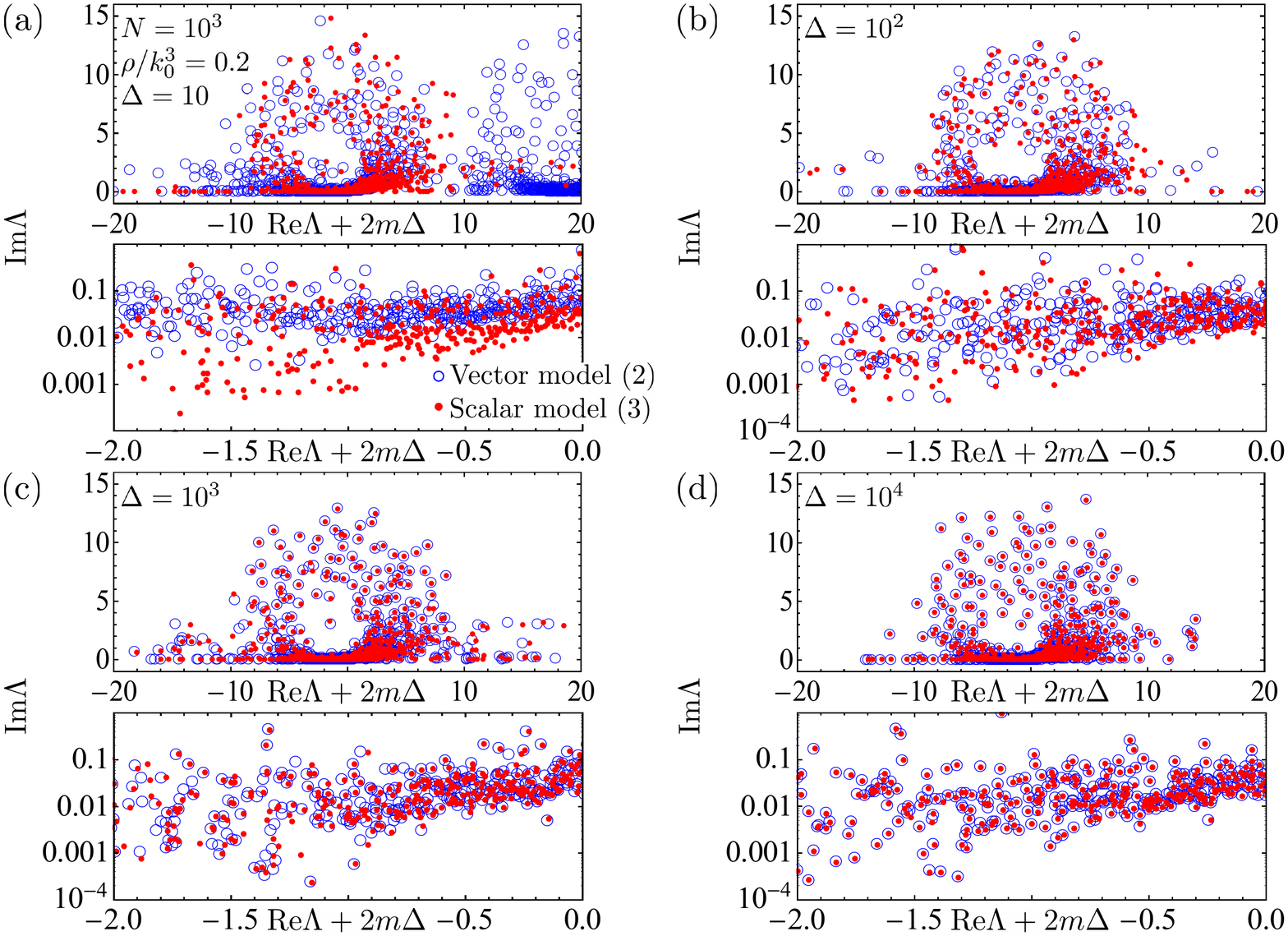}
\caption{Illustration of validity of the effective scalar model (3). For a random realization of atomic positions $\{ \vec{r}_j \}$ in a ball and a magnetic field $\Delta = 10$ (a), 100 (b), $10^3$ (c) and $10^4$ (d), the eigenvalues of the $3 N \times 3 N$ matrix $G$ defined by Eq.\ (2) (blue empty circles) are compared with the eigenvalues of the $N \times N$ matrix ${\cal G}^{(+1)}$ defined by Eq.\ (3) (red dots). For each $\Delta$, the upper panel shows the eignevalues in the linear scale whereas the lower panel is a zoom on the part of the complex plane with eigenvalues $\Lambda$ corresponding to localized eigenvectors, with a logarithmic scale for $\mathrm{Im} \Lambda$.}
\label{smfigtest}
\end{figure*}

Figure \ref{smfigtest} shows a comparison of eigenvalues of the $N \times N$ matrix ${\cal G}^{(+1)}$ defined by Eq.\ (3) (red dots) with eigenvalues of the $3N \times 3N$ matrix $G$ defined by Eq.\ (2) (blue open circles), for $N = 10^3$, $\rho/k_0^3 = 0.2$ and 4 values of magnetic field quantified by the dimensionless Zeeman shift $\Delta$. For each $\Delta$, we show the eigenvalues on the complex plane in the linear scale (upper panels) and a zoom on the part of the complex plane where the eigenvalues corresponding to spatially localized eigenvectors are located (lower panels). To better resolve eigenvalues with small imaginary parts, the logarithmic scale is used for the imaginary axis in the latter case. We see from Fig.\ \ref{smfigtest} that the eigenvalues of ${\cal G}^{(+1)}$ and $G$ having real parts $\mathrm{Re} \Lambda$ around $-2 m \Delta$ (here, for $m = 1$) indeed coincide in the limit of large $\Delta$ [e.g., for $\Delta = 10^4$, see Fig.\ \ref{smfigtest}(d)]. For a smaller $\Delta = 10^3$, the eigenvalues of the two matrices start to differ but remain quite close, so that pairs of eigenvalues of ${\cal G}^{(+1)}$ and $G$ that correspond to each other can be readily identified, see Fig.\ \ref{smfigtest}(c). At $\Delta = 10^2$ such an identification becomes difficult, but the patterns of red dots and blue circles in Fig.\ \ref{smfigtest}(b) have the same global structure, so that the statistical properties of eigenvalues are likely to be similar. And finally, at even smaller $\Delta = 10$, the eigenvalues of the two matrices start to differ substantially [see Fig.\ \ref{smfigtest}(a)]. In particular, the ``cloud'' of eigenvalues of $G$ corresponding to $m = 0$ and not described by Eq.\ (3) start to be visible in the right part of the figure. The eigenvalues of this cloud can, in their turn, be approximated by eigenvalues of an $N \times N$ matrix ${\cal G}^{(0)}$ defined by Eq.\ (\ref{g0sm}). However, adding the eigenvalues of ${\cal G}^{(0)}$ to Fig.\ \ref{smfigtest}(a) does not help to cure the main difference between the eigenvalues of the full vector model and those of its scalar approximation: the red dots in the lower panel of Fig.\ \ref{smfigtest}(a) are considerably lower than the blue circles, and we know from the previous work \cite{skip15sm,skip14sm} that this corresponds to a qualitative difference in the spatial structure of corresponding eigenvectors which are extended for the full vector model (2) at small $\Delta$ but localized in the scalar model (3) at any $\Delta$. According to Fig.\ \ref{smfigtest}, the two models start to be roughly equivalent only for $\Delta \gtrsim 10^2$.

More generally, at large $\Delta$ the three clouds of eigenvalues corresponding to $m = 0$, $\pm 1$ are roughly circular, as one can see from Fig.\ \ref{smfigtest} for the $m = 1$ cloud. If we assume that their radii ${\cal R}$ scale in the same way as for the scalar-wave model, we obtain ${\cal R} \propto b_0 \sim R/\ell_0$, where $\ell_0 \sim k_0^2/\rho$ is the on-resonance scattering mean free path computed in the ISA and $b_0$ is the optical thickness of the atomic medium \cite{skip11sm}. Obviously, replacing the $3N \times 3N$ matrix $G$ by three $N \times N$ matrices ${\cal G}^{(m)}$ is a good approximation only when the clouds of eigenvalues corresponding to different $m$ are well separated on the complex plane, i.e. when $\Delta \gg b_0$. This is the regime that we study in the main text of the paper.

\section{Localization phase diagram from the Ioffe-Regel criterion in the independent-scattering approximation}
\label{isa}

As we notice in the main text, some features of the localization phase diagram shown in Fig.\ 3 can be qualitatively understood based on the Ioffe-Regel (IR) criterion of localization evaluated in the independent-scattering approximation. The IR criterion suggests that the Anderson transition takes place when $k(\omega) \ell(\omega) \simeq 1$, where $k(\omega)$ is the effective wave number and $\ell(\omega)$ is the scattering mean free path of the wave in the medium \cite{ioffe60sm,sheng95sm}. For resonant point scatterers, the IR criterion has been recently tested in Ref.\ \cite{skip18asm}, which has shown that it can serve as a qualitative criterion of localization, whereas its quantitative validity is poor. One should not be distracted by the seemingly simple form of the IR criterion because in strongly scattering media where Anderson localization can be reached, the calculations of both $k(\omega)$ and $\ell(\omega)$ are highly nontrivial.

\begin{figure}[b]
\includegraphics[width=\columnwidth]{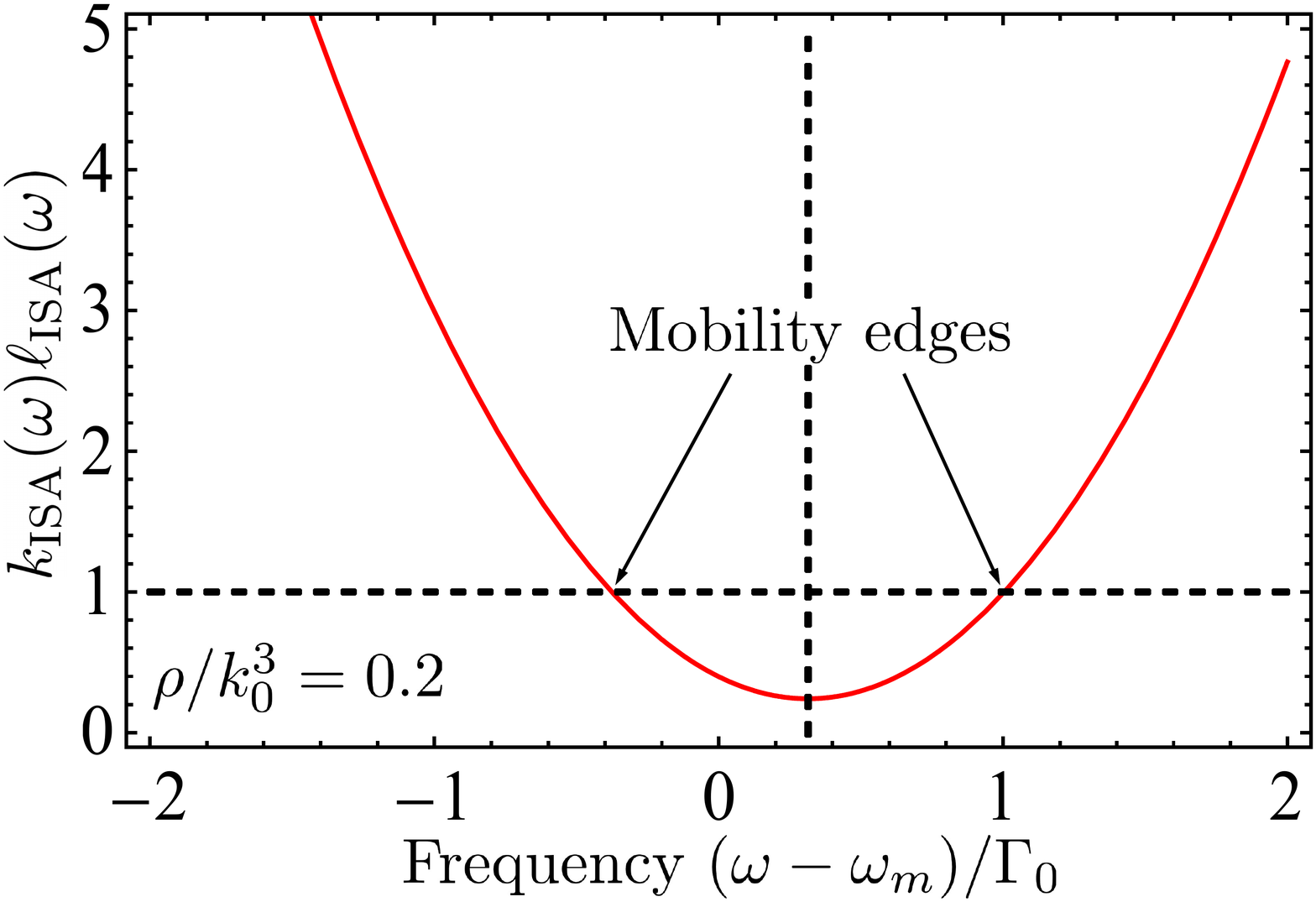}
\caption{{The Ioffe-Regel parameter $k_{\mathrm{ISA}}(\omega) \ell_{\mathrm{ISA}}(\omega)$ computed in the ISA, is shown as a function of frequency $\omega$ at a fixed density $\rho/k_0^3 = 0.2$. The horizontal dashed line shows its critical value that approximately determines the position of the mobility edges. The vertical dashed line shows the frequency for which $k_{\mathrm{ISA}}(\omega) \ell_{\mathrm{ISA}}(\omega)$ takes the minimum value.}}
\label{smfigir}
\end{figure}

A simple approximation that can be used to evaluate $k(\omega)$ and $\ell(\omega)$ in the limit of low density of scatterers is the so-called independent-scattering approximation (ISA) \cite{sheng95sm,lagendijk96sm,devries98sm}. We apply the ISA to a cloud of atoms in a strong magnetic field assuming that light is quasi-resonant with the transition $J_g =  0$ $\to$ ($J_e = 1$, $m = \pm 1$) that has a resonance frequency $\omega_m = \omega_0  + m \Gamma_0 \Delta$ and a resonance width $\Gamma_0$. In the ISA, $k-k_0$ and $1/2\ell$ are proportional to the atomic number density $\rho$ \cite{sheng95sm,lagendijk96sm,devries98sm}:
\begin{eqnarray}
k(\omega) &=& k_0  - \frac{1}{2 k_0} \mathrm{Re} \Sigma(\omega),
\label{keff}
\\
\frac{1}{2\ell(\omega)} &=& -\frac{1}{2 k_0} \mathrm{Im} \Sigma(\omega),
\label{mfp}
\\
\Sigma(\omega) &=& \Sigma_{\mathrm{ISA}}(\omega) = \rho t(\omega)
\nonumber \\
&=& \frac{4 \pi \rho}{k_0} \, \frac{1}{2(\omega - \omega_m)/\Gamma_0 + i},
\label{sigma}
\end{eqnarray}
where $k_0 = \omega/c$, we introduced the scattering matrix of an atom $t(\omega)$ and the self-energy $\Sigma(\omega)$, and assumed a scalar approximation. The latter seems to be justified because in a strong field, the atoms scatter only a single polarization component of light.
It readily follows from Eqs.\ (\ref{keff})--(\ref{sigma}) that
\begin{eqnarray}
&&k_{\mathrm{ISA}}(\omega) = k_0 \left[ 1 - \frac{4 \pi \rho}{k_0^3} \; \frac{(\omega-\omega_m)/\Gamma_0}{1 + 4(\omega-\omega_m)^2/\Gamma_0^2} \right],
\label{keff2}
\\
&&\ell_{\mathrm{ISA}}(\omega) = \frac{k_0^2}{4 \pi \rho} \left[ 1 + \frac{4(\omega-\omega_m)^2}{\Gamma_0^2} \right],
\label{mfp2}
\\
&&k_{\mathrm{ISA}}(\omega) \ell_{\mathrm{ISA}}(\omega) = \frac{k_0^3}{4 \pi \rho} \left[ 1 + \frac{4(\omega - \omega_m)^2}{\Gamma_0^2} \right]
\nonumber \\
&&\hphantom{k_{\mathrm{ISA}}(\omega) \ell_{\mathrm{ISA}}(\omega)} - \frac{\omega - \omega_m}{\Gamma_0}.
\label{irkl}
\end{eqnarray}

Even though Eqs.\ (\ref{keff})--(\ref{irkl}) are valid only at small $\rho$, we can still hope that they yield physically reasonable (even though quantitatively inaccurate) results when $\rho$ is large and the IR criterion of localization is satisfied.
{The frequency dependence of $k_{\mathrm{ISA}}(\omega) \ell_{\mathrm{ISA}}(\omega)$ is shown in Fig.\ \ref{smfigir} for $\rho/k_0^3 = 0.2$.}
Requiring $k_{\mathrm{ISA}}(\omega) \ell_{\mathrm{ISA}}(\omega) \leq 1$ leads to the following results. First, $k_{\mathrm{ISA}}(\omega) \ell_{\mathrm{ISA}}(\omega) \leq 1$ can only be reached for densities $\rho/k_0^3 > \rho_c/k_0^3 = (\sqrt{5} - 2)/\pi \simeq 0.075$, which is close to $\rho_c/k_0^3 \simeq 0.1$ found from the calculations in the main text of the paper. Second,
{the minimum of $k_{\mathrm{ISA}}(\omega) \ell_{\mathrm{ISA}}(\omega)$ is reached at $\omega = \omega_m + \Gamma_0 \pi\rho/2 k_0^3$, and hence}
the frequency $\omega_c$ at which the condition $k_{\mathrm{ISA}}(\omega) \ell_{\mathrm{ISA}}(\omega) = 1$ is first obeyed, is blue shifted with respect to $\omega_m$: $(\omega_c - \omega_m)/\Gamma_0 = \sqrt{5}/2 - 1 \simeq 0.12 > 0$. This is in qualitative agreement with our numerical results presented in Fig.\ 3, even though $(\omega_c - \omega_m)/\Gamma_0 \simeq 0.5$ found from the latter is significantly larger. The blue shift of $\omega_c$ can be understood from Eq.\ (\ref{keff2}) that shows that $k < k_0$ on the right from the resonance ($\omega > \omega_m$), whereas $k > k_0$ on the left from it ($\omega < \omega_m$). As a consequence, the {the Ioffe-Regel criterion is easier to obey} on the right from the resonance.
{Note that neither $\omega_c$ nor the frequency at which the minimum of  $k_{\mathrm{ISA}}(\omega) \ell_{\mathrm{ISA}}(\omega)$ is reached in Fig.\ \ref{smfigir} correspond to the position of the atomic resonance in the system. The latter can be defined as a position of either the maximum of fluorescence spectrum or the minimum of continuous-wave transmission and may be subject to the Lorentz-Lorenz local field or collective Lamb shifts. The latter shifts were recently shown to be absent for ensembles of cold (i.e., motionless atoms) that we study here, and require inhomogeneous broadening (e.g., Doppler effect) or some other effect that would suppress correlations between atoms \cite{jav14,jen16}.
}

 Finally, for $\rho > \rho_c$, two mobility edges follow from Eq.\ (\ref{irkl}):
\begin{eqnarray}
\frac{\omega_c - \omega_m}{\Gamma_0} &=& \frac{\pi}{2 k_0^3} \left[ \rho
\pm \sqrt{\left(  \rho - \rho_c  \right)
\left(  \rho + \rho^*  \right)}
\right].
\label{meisa}
\end{eqnarray}
where we defined $\rho^* = k_0^3 (\sqrt{5} + 2)/\pi \simeq 1.35 k_0^3$.
The width of the frequency band of localized states readily follows:
\begin{eqnarray}
&&\frac{\Delta \omega}{\Gamma_0} = \frac{\pi}{k_0^3} \sqrt{\left(  \rho - \rho_c  \right)
\left(  \rho + \rho^*  \right)}.
\label{locwidth}
\end{eqnarray}
These results are shown by dash-dotted lines in Fig.\ 3 and turn out to provide a qualitatively correct picture of the localization phase diagram.

The above calculation may be modified and extended in several ways. First, the vector nature of light can be taken into account by replacing the numerical factors $4\pi$ in Eqs.\ (\ref{sigma})--(\ref{irkl}) by $6\pi$ \cite{devries98sm}. Second, one can use a critical value $(k\ell)_c$ of the IR parameter $k\ell$ that is different from 1. All these modifications do not considerably improve the overall agreement of the final results for the location of mobility edges and the width of the spectral band of localized states with our numerical results in Fig.~3.

\end{document}